# Trial of a search for a face-centered-cubic high-entropy alloy superconductor

**Naoki Ishizu** [1] **and Jiro Kitagawa** [1]

[1] Department of Electrical Engineering, Faculty of Engineering, Fukuoka Institute of Technology, 3-30-1 Wajiro-higashi, Higashi-ku, Fukuoka 811-0295, Japan

**Abstract:** With the aim of the discovery of face-centered-cubic (fcc) high-entropy alloy (HEA) superconductor, we have carried out materials research on Nb or Pb-containing multi-component alloys. Although the X-ray diffraction (XRD) patterns of some Nb-containing samples exhibited the dominant fcc phases, no superconducting signals were observed down to 3 K. Examination with an energy dispersive X-ray spectrometer revealed that all samples were multi-phase, but the existence of several new Nb-containing HEA phases was found in them. It was confirmed that the synthesis of Pb-containing an HEA or quaternary alloy would be difficult, probably due to the large differences in the crystal structure and atomic radius among constituent elements, low reaction temperature and the lack of a rapid cooling process in the synthesis. Despite the negative results in this research, some hints for an improved strategy for the search for an fcc HEA superconductor are provided. Moreover, our results are useful as fundamental data for future HEA predictions or for studies of phase relations in Nb or Pb-containing multi-component alloys based on the CALPHAD (calculation of phase diagram) method.

## 1. Introduction

High-entropy alloys (HEAs) have attracted a great deal of attention as a new class of materials [1-4]. A typical HEA forms a face-centered-cubic (fcc), body-centered-cubic (bcc) or hexagonal-closed packing (hcp) structure. The most prominent feature of an HEA is that more than five elements, with the atomic fraction of each element being between 5% and 35%, randomly occupy one crystallographic site. This leads to a large mixing entropy, and HEAs show the combination of a high yield strength and ductility [5], high strength at high temperatures [6], strong resistance to corrosion and oxidation [7] and so on.

HEA superconductivity is a very new research topic which has been investigated since 2014. Transition metal-based superconductors with simple crystalline structures follow the so-called Matthias rule, showing broad peak structures at the specified valence electron count (VEC) per atom when the superconducting critical temperature $T_c$ is plotted as a function of VEC per atom [8]. On the other hand, transition metal amorphous superconductors do not follow this rule and frequently show rather high $T_c$ values in the valley of the curve of the Matthias rule [9]. HEA superconductors with simple crystal structures have been found in bcc [10-17] and hcp [18-20] forms, the $T_c$ vs VEC plots of which seem to fall between a crystalline curve and an amorphous one. Thus, it is expected that HEA superconductors will be useful for the study of the relationship between crystalline and amorphous compounds.

The excellent review in [21] of HEA superconductors proposed a possible fcc HEA superconductor. While it has been recently discovered that fcc-related NaCl-type HEAs show superconducting states [22, 23], a simple fcc HEA superconductor has not yet been discovered. An fcc HEA superconductor would contribute to the deep understanding of HEA and/or of the relationship between crystalline and amorphous compounds. In this paper, we introduce our attempt at the search for an fcc HEA superconductor. Our strategy is the employment of a rather high $T_c$ element, because the HEA superconductors reported to date contain superconducting elements. We focused on the Nb- and Pb- containing HEAs.

## 2. Materials and Methods

Nb-containing samples were synthesized by a home-made arc furnace. The constituent elements were arc-melted on a water-cooled Cu hearth in an Ar atmosphere. For a Pb-containing sample, the constituent elements were sealed in an evacuated quartz tube, which was then heated in an electric furnace. The materials used are listed in Table 1 with the supply companies and the purities.

A powder X-ray diffractometer (XRD-7000L, Shimadzu, Kyoto, Japan) with Cu-K$\alpha$ radiation was employed to detect the X-ray diffraction (XRD) patterns of prepared samples. The 2$\theta$ range was between 10° and 90°. The microstructure of each sample was checked by a field emission scanning electron microscope (FE-SEM, JSM-7100F; JEOL, Akishima, Japan). The atomic compositions of the samples were determined by an energy dispersive X-ray (EDX) spectrometer equipped to the FE-SEM.

In order to confirm the diamagnetic signal due to the superconducting state, the temperature dependence of the AC magnetic susceptibility $\chi_{ac}$ ($T$), in which the amplitude and the frequency of the AC field were 5 Oe and 800 Hz, respectively, was checked by a home-made system in a GM refrigerator (UW404, Ulvac cryogenics, Kyoto, Japan) between 3 and 300 K.

**Table 1.** Materials used in this study. The supply company, purity, crystal structure at room temperature, atomic radius [3] and valence electron count (VEC) are also listed.

| Element | Supply Company | Purity (%) | Crystal structure | Atomic radius (Å) | VEC |
|---|---|---|---|---|---|
| Zr | Soekawa Chemicals, Tokyo, Japan | 99 | A3 (hcp) | 1.6025 | 4 |
| Nb | Nilaco, Tokyo, Japan | 99.9 | A2 (bcc) | 1.429 | 5 |
| V | Kojundo Chemical Laboratory, Sakado, Japan | 99.9 | A2 (bcc) | 1.316 | 5 |
| Ru | Soekawa Chemicals, Tokyo, Japan | 99.9 | A3 (hcp) | 1.3384 | 8 |
| Ir | Furuya Metal, Tokyo, Japan | 99.99 | A1 (fcc) | 1.3573 | 9 |
| Rh | Soekawa Chemicals, Tokyo, Japan | 99.9 | A1 (fcc) | 1.345 | 9 |
| Pd | Tanaka Kinzoku Kogyo, Tokyo, Japan | 99.9 | A1 (fcc) | 1.3754 | 10 |
| Cu | Soekawa Chemicals, Tokyo, Japan | 99.99 | A1 (fcc) | 1.278 | 11 |
| Pb | Osaka Asahi Metal, Osaka, Japan | 99.9999 | A1 (fcc) | 1.7497 | 4 |
| Sn | Kojundo Chemical Laboratory, Sakado, Japan | 99.99 | A5 | 1.62 | 4 |
| Bi | Osaka Asahi Metal, Osaka, Japan | 99.9999 | A7 | 1.6 | 5 |
| Sb | Soekawa Chemicals, Tokyo, Japan | 99.999 | A7 | 1.45 | 5 |
| Te | Kojundo Chemical Laboratory, Sakado, Japan | 99.9999 | A8 | 1.452 | 6 |

## 3. Results and Discussion

### 3.1. Nb-containing samples

The starting compositions of prepared Nb-containing samples were basically determined following the conventional design rule [1]: a $\delta$-value less than 5 % and a VEC larger than 8.0. In order to realize the requirements, fcc elements were predominantly employed (see also Table 1 and Table 2). $\delta$ and VEC are calculated as follows:

$$\delta = 100 \times \sqrt{\sum_{i=1}^{n} c_i \left(1 - \frac{r_i}{\bar{r}}\right)^2} \tag{1}$$

and

$$\text{VEC} = \sum_{i=1}^{n} c_i \text{VEC}_i, \tag{2}$$

where $c_i$, $r_i$ and $\text{VEC}_i$ are the atomic fraction, the atomic radius and the VEC of element $i$, respectively, and $\bar{r}$ is the composition-weighted average atomic radius. The parameter $\delta$ is a measure of the degree of the atomic size difference among the constituent elements. These parameters are calculated for the prepared samples and are listed in Table 2, in which the samples are named as their starting compositions. $Nb_{15}V_{10}Rh_{30}Pd_{25}Cu_{20}$, $Nb_{15}Ir_{21}Rh_{21}Pd_{22}Cu_{21}$ and $Nb_{17}Ru_{12}Ir_{10}Rh_{28}Pd_{33}$ fulfill the design requirements. To evaluate the effect of larger $\delta$ values, two samples including a Zr atom are prepared.

**Table 2.** $\delta$ and VEC of Nb-containing samples and phases detected by energy dispersive X-ray (EDX) measurements.

| No. | Sample | Composition of Phase I, II or III | $\delta$ | VEC |
|---|---|---|---|---|
| 1 | $Nb_{15}V_{10}Rh_{30}Pd_{25}Cu_{20}$ | | 3.52 | 8.65 |
| | Phase I | $Nb_{21.2(8)}V_{6.0(5)}Rh_{42.9(5)}Pd_{21.6(4)}Cu_{8.3(8)}$ | 3.14 | 8.29 |
| | Phase II | $Nb_{12.6(5)}V_{18.6(5)}Rh_{26.4.(8)}Pd_{28(1)}Cu_{14.4(5)}$ | 3.26 | 8.32 |
| | Phase III | $Nb_{2(1)}V_{3(1)}Rh_{2(1)}Pd_{28(2)}Cu_{65(5)}$ | 3.56 | 10.38 |
| 2 | $Nb_{15}Ir_{21}Rh_{21}Pd_{22}Cu_{21}$ | | 3.45 | 9.04 |
| | Phase I | $Nb_{24.3(4)}Ir_{36.4(8)}Rh_{27.8(3)}Pd_{10.1(9)}Cu_{1.4(6)}$ | 2.52 | 8.16 |
| | Phase II | $Nb_{12.5(6)}Ir_{6(1)}Rh_{27(2)}Pd_{46(3)}Cu_{8.5(7)}$ | 2.66 | 9.13 |
| | Phase III | $Nb_{4(1)}Rh_{4(1)}Pd_{52(3)}Cu_{40(5)}$ | 3.74 | 10.16 |
| 3 | $Zr_{21}Nb_{15}Rh_{21}Pd_{22}Cu_{21}$ | | 8.00 | 7.99 |
| | Phase I | $Zr_{24.3(5)}Rh_{19.2(4)}Pd_{37.2(2)}Cu_{19.3(3)}$ | 8.30 | 8.54 |
| | Phase II | $Zr_{17(1)}Nb_{41(1)}Rh_{27(1)}Pd_{8(1)}Cu_{7(1)}$ | 6.60 | 6.73 |
| | Phase III | $Zr_{21(1)}Rh_{9(1)}Pd_{13(1)}Cu_{57(1)}$ | 9.33 | 9.22 |
| 4 | $Zr_6Nb_{15}V_{10}Rh_{25}Pd_{24}Cu_{20}$ | | 5.62 | 8.34 |
| | Phase I | $Zr_{8.7(5)}Nb_{15.7(5)}V_{7(1)}Rh_{23.1(5)}Pd_{31(1)}Cu_{14.5(3)}$ | 5.98 | 8.26 |
| | Phase II | $Nb_{24(1)}V_{15.8(8)}Rh_{27.2(5)}Pd_{18(1)}Cu_{15.0(5)}$ | 3.72 | 7.89 |
| | Phase III | $Pd_{14(1)}Cu_{86(1)}$ | 2.62 | 10.86 |
| 5 | $Nb_{20}V_{10}Pd_{30}Cu_{40}$ | | 4.44 | 8.9 |
| | Phase I | $Nb_{27.6(2)}V_{11.5(7)}Pd_{39.7(7)}Cu_{21.1(2)}$ | 4.05 | 8.25 |
| | Phase II | $Pd_{10.8(5)}Cu_{89.2(5)}$ | 2.35 | 10.89 |
| 6 | $Nb_{17}Ru_{12}Ir_{10}Rh_{28}Pd_{33}$ | | 2.21 | 8.54 |
| | Phase I | $Nb_{17.7(8)}Ru_{16.8(5)}Ir_{15.5(2)}Rh_{29.7(2)}Pd_{20.3(7)}$ | 2.30 | 8.33 |
| | Phase II | $Nb_{16.0(5)}Ru_{7.0(7)}Ir_{4.0(6)}Rh_{22(1)}Pd_{51(1)}$ | 2.02 | 8.8 |

Figure 1 shows the XRD patterns of prepared samples. In the upper five samples, which all contain Nb, Pd and Cu atoms, $Nb_{15}V_{10}Rh_{30}Pd_{25}Cu_{20}$ and $Nb_{15}Ir_{21}Rh_{21}Pd_{22}Cu_{21}$ show dominant fcc phases. On the other hand, the XRD patterns of $Zr_{21}Nb_{15}Rh_{21}Pd_{22}Cu_{21}$ and $Zr_6Nb_{15}V_{10}Rh_{25}Pd_{24}Cu_{20}$ cannot be characterized by fcc phases. The comparison of starting compositions, exhibiting contrasting results, suggests that Zr is unfavorable for the formation of an fcc structure. To further elucidate the formation condition of the single fcc phase, the quaternary alloy $Nb_{20}V_{10}Pd_{30}Cu_{40}$ was synthesized. As shown in Fig. 1, this sample shows two fcc phases with quite different lattice parameters. The XRD pattern of the sample with no Cu atom (see the bottom of Fig. 1) can be explained by an fcc phase. The lattice parameters of all fcc phases were obtained by the least-square method [24, 25] and are exhibited in Fig. 1. The fitting range of $2\theta$ was between 10° and 90°. The SEM images of all samples are displayed in Figs. 2(a) to 2(f), suggesting that all samples are multi-phase. In $Nb_{15}V_{10}Rh_{30}Pd_{25}Cu_{20}$ (Fig. 2(a)) and $Nb_{15}Ir_{21}Rh_{21}Pd_{22}Cu_{21}$ (Fig. 2(b)), three contrast phases I, II and III were observed. In each case, the brightest area (phase I) showed a dendritic morphology. Phase II with the median contrast seems to surround phase I, and the darkest area (phase III) is the precipitate which formed in the final solidification process. $Zr_{21}Nb_{15}Rh_{21}Pd_{22}Cu_{21}$ (Fig. 2(c)) or $Zr_6Nb_{15}V_{10}Rh_{25}Pd_{24}Cu_{20}$ (Fig. 2(d)) partially showed a eutectic-like structure formed by phase I and phase II (see the green elliptic closed-curve). As shown in Fig. 2(e), $Nb_{20}V_{10}Pd_{30}Cu_{40}$ possessed two phases, both of which were fcc phases. $Nb_{17}Ru_{12}Ir_{10}Rh_{28}Pd_{33}$ possessed two contrast areas (see phases I and II in Fig. 2(f)). The shape of the main phase had a dendritic-like morphology. The compositions of all phases determined by EDX are listed in Table 2.

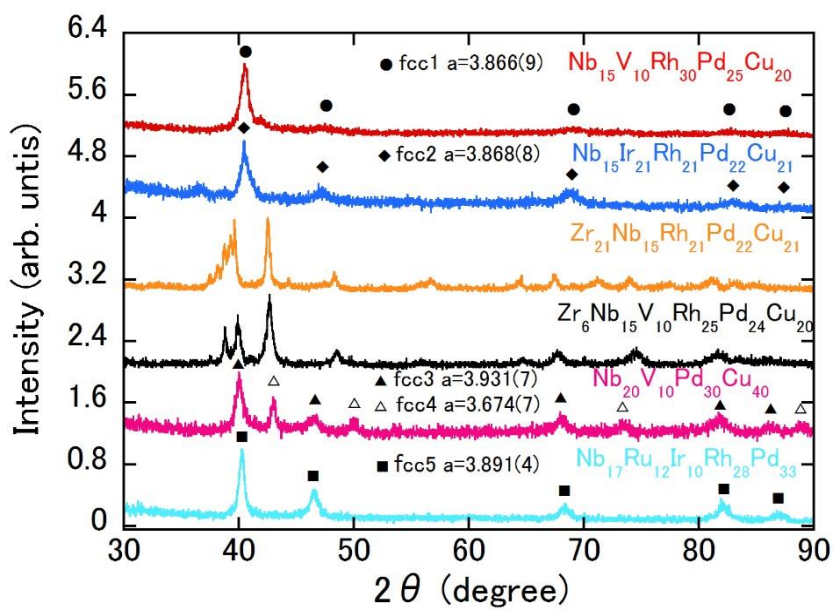

**Figure 1.** X-ray diffraction (XRD) patterns of Nb-containing samples. The origin of each pattern is shifted by an integer value.

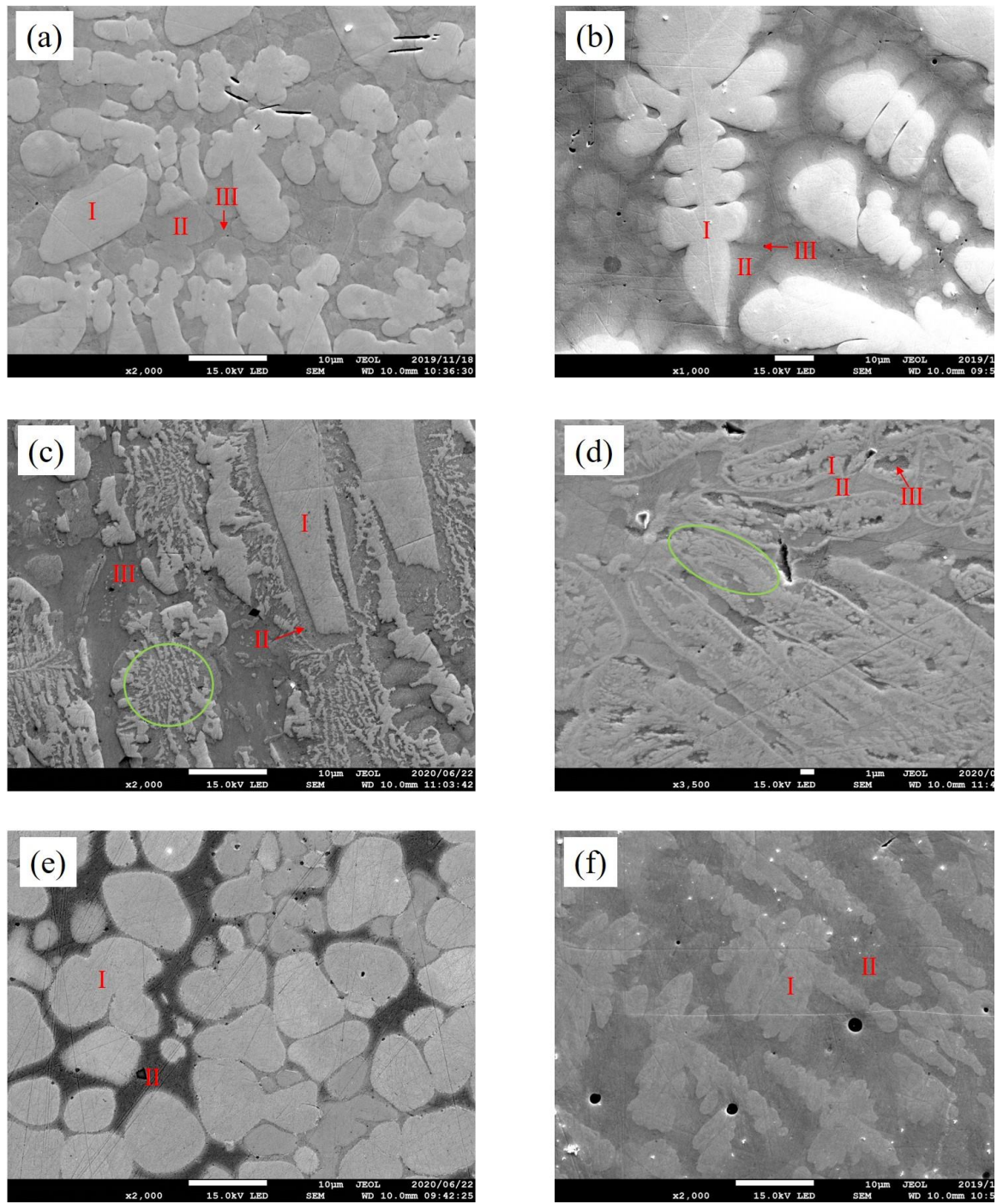

**Figure 2.** Back-scattered electron (15keV) images of (a) $Nb_{15}V_{10}Rh_{30}Pd_{25}Cu_{20}$, (b) $Nb_{15}Ir_{21}Rh_{21}Pd_{22}Cu_{21}$, (c) $Zr_{21}Nb_{15}Rh_{21}Pd_{22}Cu_{21}$, (d) $Zr_6Nb_{15}V_{10}Rh_{25}Pd_{24}Cu_{20}$, (e) $Nb_{20}V_{10}Pd_{30}Cu_{40}$ and (f) $Nb_{17}Ru_{12}Ir_{10}Rh_{28}Pd_{33}$, respectively.

All prepared samples were examined by $\chi_{ac}(T)$ measurements, which detected no superconducting signals down to 3 K. These results suggest that an Nb-containing fcc-HEA might be an inadequate strategy for searching for fcc HEA superconductors. At the present stage, we speculate that the appearance of superconductivity in a Nb-containing fcc compound is a rare event because almost all Nb-based superconductors form bcc-related structures. The rare example is NbN or NbC, crystallizing into an NaCl-type structure, which is related to the fcc structure [26, 27]. While our results are negative for the research of Nb-containing fcc HEA superconductors, it is valuable to report that phase II in sample no. 1, phase I in sample no. 6 and possibly phases I and II in sample no. 4 are new members of HEA.

Here, we discuss the fcc phase stability, viewed from the perspective of $\delta$ and VEC, which is summarized in Table 2. We also calculate the parameters for the phases detected by EDX. The values of $\delta$ were very large in $Zr_{21}Nb_{15}Rh_{21}Pd_{22}Cu_{21}$ and $Zr_6Nb_{15}V_{10}Rh_{25}Pd_{24}Cu_{20}$ due to the larger atomic radius of Zr, and this caused there to be no fcc phase in each sample. While $\delta$ was reduced in $Nb_{15}V_{10}Rh_{30}Pd_{25}Cu_{20}$ and $Nb_{15}Ir_{21}Rh_{21}Pd_{22}Cu_{21}$, each sample possessed three phases. In each case, moving from phase I to phase III with a rather low $\delta$ for the respective phases, the VEC value increased, which accompanied the decrease (increase) of the Nb (Cu) atomic fraction. This suggests that the combination of Nb and Cu is not recommended even with a reduced $\delta$, because an Nb-rich phase and a Cu-rich phase are stabilized for a smaller VEC and a larger VEC, respectively. Probably due to this reason, quaternary $Nb_{20}V_{10}Pd_{30}Cu_{40}$ does not exhibit a single fcc phase. $Nb_{17}Ru_{12}Ir_{10}Rh_{28}Pd_{33}$ with substantially suppressed $\delta$ and no Cu atom was expected to show a single fcc phase; nonetheless, two phases were detected in the sample. The detected phases possessed reduced $\delta$ values and a similar VEC. Thus, it may be inherently difficult to obtain a single-phase Nb-containing fcc HEA.

*3.2. Pb-containing samples*

As shown in Table 1, the atomic radius of fcc Pb is much bigger than those of the fcc elements Ir, Rh, Pd et al. Moreover, fcc elements do not exist near to Pb in the periodic table. Nonetheless, we tentatively employed Pb and the neighboring elements (Sn, Bi, Sb and Te) in the periodic table according to the HEA makeup shown in Table 3. Figure 3 shows the XRD patterns of prepared samples, which are described by their starting compositions. The reaction temperatures were 700 ℃ for PbSnBiSbTe, 800 ℃ for PbBiSbTe and 850 ℃ for PbSnBiSb, respectively. The XRD pattern of PbSnBiSbTe was mainly the same as that of PbTe. As mentioned below, PbSnBiSbTe contained no phase composed of five elements; therefore, we feel that a Pb-containing HEA is difficult to synthesize, and have changed our aim to the search for a quaternary Pb-containing alloy. Thus, the prepared PbBiSbTe and PbSnBiSb showed XRD patterns with assigned phases of PbTe and PbBi + SnSb + Bi, respectively. The SEM images of all samples are shown in Figs. 4(a) and 4(c). All samples possessed multi-phases, as denoted in the figures. The determined compositions of these phases are listed in Table 3 and are consistent with the phase assignments in XRD patterns.

**Table 3.** $\delta$ and VEC of Pb-containing samples and phases detected by EDX measurements.

| No. | Sample | Composition of Phase I, II, III or IV | $\delta$ | VEC |
|---|---|---|---|---|
| 7 | PbSnBiSbTe | | 7.18 | 4.8 |
| | Phase I | $Pb_{34.3(4)}Sn_{17.7(2)}Te_{48.0(4)}$ | | |
| | Phase II | $Bi_{95(5)}Sb_{5(5)}$ | | |
| | Phase III | $Pb_{1.8(8)}Sn_{48.6(3)}Bi_{5.2(5)}Sb_{44.4(9)}$ | | |
| | Phase IV | $Pb_{48(1)}Sn_{7(3)}Bi_{45(1)}$ | | |
| 8 | PbBiSbTe | | 7.92 | 5 |
| | Phase I | $Pb_{51.1(2)}Te_{48.9(2)}$ | | |
| | Phase II | $Bi_{44.3(5)}Sb_{55.7(5)}$ | | |
| 9 | PbSnBiSb | | 6.62 | 4.5 |
| | Phase I | $Pb_{3.0(3)}Sn_{46(1)}Bi_{3.0(1)}Sb_{48(1)}$ | | |
| | Phase II | $Pb_{55(1)}Sn_{4(1)}Bi_{41(1)}$ | | |
| | Phase III | $Sn_{2(1)}Bi_{95(1)}Sb_{3(1)}$ | | |

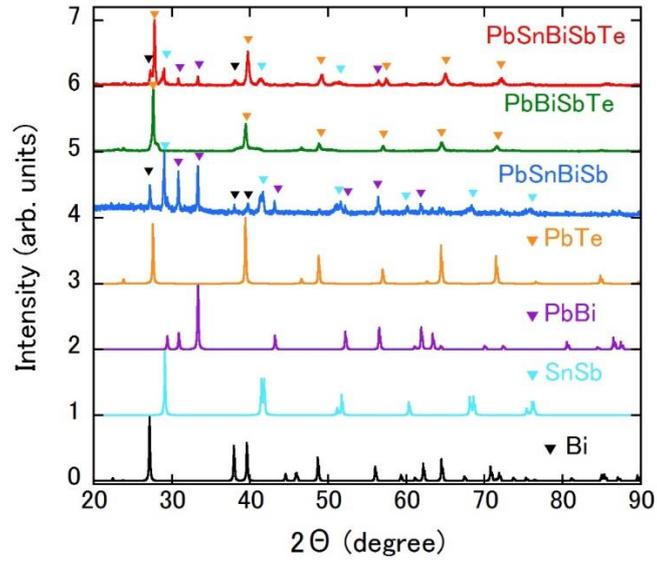

**Figure 3.** XRD patterns of Pb-containing samples. The simulated patterns of PbTe, PbBi, SnSb and Bi are also shown. The origin of each pattern is shifted by an integer value.

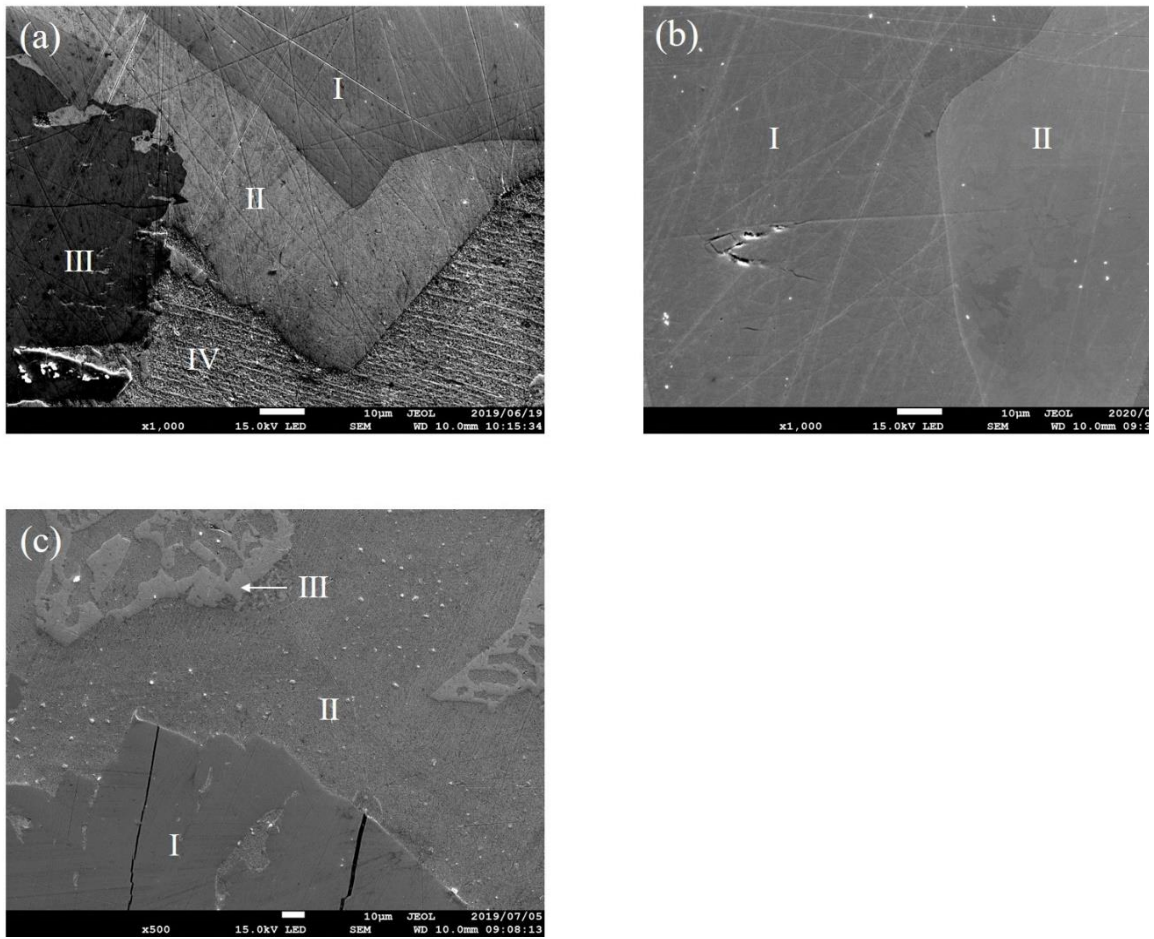

**Figure 4.** Back-scattered electron (15 keV) images of (a) PbSnBiSbTe, (b) PbBiSbTe and (c) PbSnBiSb, respectively.

As shown in Fig. 5, PbSnBiSbTe and PbSnBiSb exhibit diamagnetic signals at approximately 8.6 K, while PbBiSbTe shows a maintained normal state down to 3 K. The diamagnetic signals are ascribed to the superconducting state of the PbBi alloy [28]. We expect that a Pb-containing fcc HEA would be promising as a superconductor if it exists because the bcc and hcp HEA superconductors reported to date contain superconducting elements and Pb is a rather high $T_c$ element. However, the synthesized samples did not contain alloys composed of four or five elements with respective atomic fractions between 5 % and 35 %, in contrast to the Nb-containing samples. The major reasons for this may be the large difference in the crystal structure of Pb (fcc) and the other elements (see Table 1) and the larger $\delta$ (see Table 3). Other speculative reasons are the low reaction temperature and the lack of a rapid cooling process in the synthesis process. According to the thermal dynamics theory, alloy formation is determined by the Gibbs free energy of mixing [1],

$$\Delta G_{\text{mix}} = \Delta H_{\text{mix}} - T\Delta S_{\text{mix}}, \qquad (3)$$

where $\Delta H_{\text{mix}}$ and $\Delta S_{\text{mix}}$ are the mixing enthalpy and the mixing entropy, respectively, and $T$ is the temperature. While a negative $\Delta H_{\text{mix}}$ stabilizes the formation of intermetallic compounds, a large value of $T\Delta S_{\text{mix}}$ exceeding the absolute value of $\Delta H_{\text{mix}}$ favors a homogeneous mixing of components. Therefore, the high mixing entropy tends to allow HEA formation, especially under a high-temperature condition. In the arc-melting process, high-temperature melting and rapid cooling after the melting through a water-cooled hearth can be realized, and Nb-containing HEA phases can be obtained. On the other hand, the rather low reaction temperature and lack of rapid cooling represents insufficient synthesis conditions for Pb-containing multi-component alloys, although a high melting temperature is difficult to realize due to the high vapor pressure of Pb.

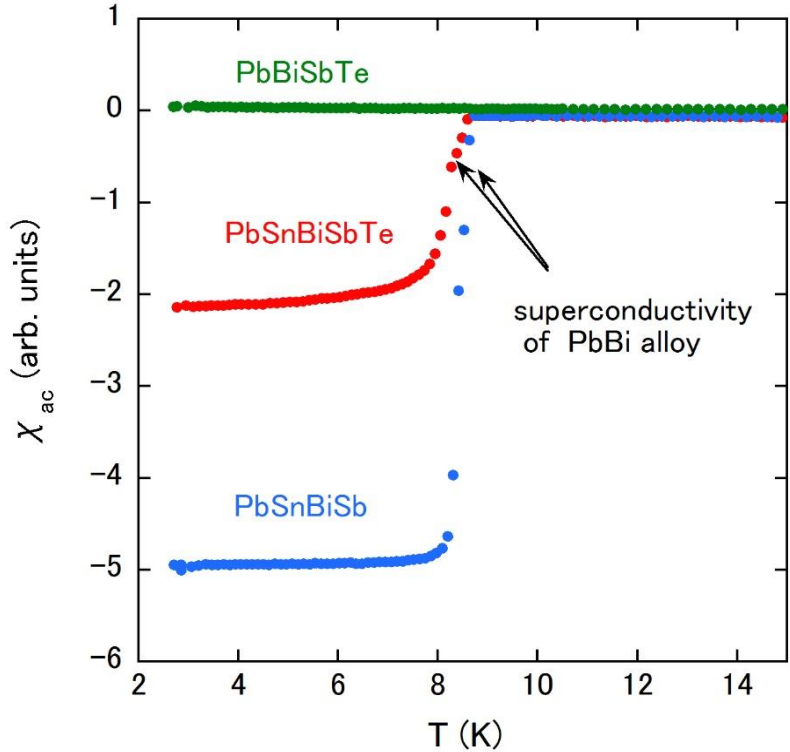

**Figure 5.** Temperature dependencies of $\chi_{\text{ac}}$ for Pb-containing samples.

## 4. Summary

We have carried out materials research into fcc HEA superconductors based on two strategies, taking into account the fact that the inclusion of rather high $T_c$ elements is advantageous. One strategy was the use of an Nb-containing HEA and the other used a Pb-containing HEA. In the studies of Nb-containing HEAs, some samples showed dominant fcc phases; however, single-phase samples could not be obtained. While we have found several new Nb-containing HEA phases in the multi-phase samples, no superconducting signal appeared in each HEA phase down to 3 K. Considering that Nb forms a bcc structure at room temperature and there are only a few examples of fcc-related Nb-based superconductors, the discovery of Nb-containing fcc HEA superconductors would be a rare event. In Pb-containing samples, even a quaternary alloy is difficult to obtain, which is partially due to the large difference in the crystal structure between Pb and the other elements and the larger $\delta$. The low reaction temperature and lack of a rapid cooling process in the synthesis might also contribute to the negative results. Our study presents some possibilities to other researchers pursuing an fcc superconducting HEA. In the research area of HEAs, the CALPHAD (calculation of phase diagram) method is rapidly used for the prediction of HEAs or the study of the phase relation between HEAs and other alloys. If the thermodynamic data of various compounds in the present Nb or Pb-containing multi-component systems are sufficiently collected, the CALPHAD method will be able to elucidate the stability of an HEA in each system. Thus, our results will greatly assist in the evaluation of the CALPHAD method in the future.